# CdSe-Au nanorod networks welded by gold domains - a promising structure for nano-optoelectronic components

Peigang Li, Alexandros Lappas*

*Institute of Electronic Structure & Laser, Foundation for Research and Technology-Hellas, Vassilika Vouton, 71110 Heraklion, Greece*

Romain Lavieville, Yang Zhang, Roman Krahne[+]

*Nano fabrication unit, Instituto Italiano di Technologia, via Morego 30, 16163 Genoa, Italy*

Abstract:

CdSe-Au networks were synthesized by a colloidal chemistry technique. They entail CdSe nanorods with a diameter of ~10 nm and a length of ~40 nm, which are joined together by Au domains (~5 nm). Individual networks were positioned by AC dielectrophoresis between bow-tie electrodes with a gap of ~100 nm and their conductivity as well as the photoelectrical properties was investigated. Nanorod networks, with multiple Au domains on the nanorod surface, displayed stable conductivity that was not sensitive to blue laser light illumination. Such nanostructures were transformed by thermal annealing to networks with Au domains only at the nanorod tips. In this system the overall conductivity was reduced, but a clear photocurrent signal could be detected, manifesting semiconductor behavior.

Introduction

In the last two decades the electrical properties of single, shape-controlled semiconductor colloidal nanocrystals have attracted much attention, especially regarding their potential applications in nanoelectronics and nanophotonics (Klein et al. 1997; Cui et al. 2005; Trudeau et al. 2008; Sheldon et al. 2009; Teich-McGoldrick et al. 2009;

*email: lappas@iesl.forth.gr
[+]email: Roman.Krahne@iit.it

Steinberg et al. 2009; Steinberg et al. 2010). For example, single-electron transistor functionality has been demonstrated based on CdSe nanoparticles and CdTe tetrapods (Klein et al. 1997; Cui et al. 2005). More recently, CdSe nanorods (NRs) have been studied for their applicability in transistor structures and for electrical current switching (Trudeau et al. 2008; Steinberg et al. 2010), demonstrating their usefulness in the development and building of nanoscale components in modern electronics. Regarding small-scale ensembles of nanocrystals, the electrical transport is hindered by the high tunnel barriers imposed by the surfactant molecules (typically phosphoric or oleic acids) that are needed to passivate the nanocrystal surface (Fig. 1a-b). Recently, Figuerola et al. have successfully assembled Au-tipped CdSe NRs end-to-end via a nanowelding approach mediated by Au domains into 1D, 2D or 3D networks (Fig. 1c)( Figuerola et al. 2008). Such all-inorganic networks, in which the NRs are linked by metal domains, promise to be an interesting alternative to overcome the problem of the high tunnel barriers (Fig. 1d). Concerning the photoelectrical properties of NRs, so far mostly systems consisting of a very large number of nanoparticles have been investigated, like in thin films deposited on micrometer spaced electrodes (Steiner et al. 2009; Persano et al. 2004). Little has been done towards the photoconductive properties of single semiconductor nanoparticles or of small-scale ensembles of those. This is due to the difficulties in fabricating electrical contacts, and to the small optical cross section of few nanoparticles that makes detectable generation of photo-induced current hard. In this respect submicron NR networks represent an interesting intermediate system. On one hand the conductive path across many NRs is facilitated by the all-inorganic interfaces within the network. On the other hand such networks provide a sufficient optical cross section that could make the observation of photocurrent possible.

In this communication, we report the implementation of CdSe-Au networks, synthesized according to the procedure published by Figuerola et al. (Figuerola et al. 2009), into bow-tie electrode devices, and the study of their conductive and photo-electrical properties. The networks were assembled in between electrodes with 100 nm gap size by AC dielectrophoresis. Scanning electron microscopy inspection confirmed the trapping of single network structures in our devices with a success rate of ∼ 85%. Electrical characterization showed stable and reproducible conductive behavior for the networks, and photo-induced conductivity in the case of networks that have been annealed.

**Experimental**

The starting CdSe NRs were prepared by colloidal chemistry via the seeded growth approach (Carbone et al. 2007). The gold domains were then selectively grown on to the tips of CdSe NRs to form a dumbbell structure (Mokari et al. 2004). In order to link the nanorods into a network, a suitable quantity of iodine ($I_2$) solution was added to the nanodumbbell solution (Figuerola et al. 2009). Samples of the network solution deposited on carbon-coated grids were taken for transmission electron microscopy (TEM) inspection (JEM 1011, JEOL Company, Japan).

Electron beam lithography (EBL) in combination with metal deposition of 5 nm Ti and 50 nm Au, followed by a standard lift-off process, was used to fabricate the bow-tie electrodes on Si substrate with a 100 nm thick oxidized layer on the surface. A typical device chip consisted of 10 pairs of electrodes with an average gap size of 100 nm.

For the network positioning via AC dielectrophoresis we used a cavity with dimensions of 1 cm × 1 cm × 0.3 cm (L×W×H) that was manufactured into a Teflon block. Prior to the trapping, the chip was cleaned successively first with acetone, then with methanol and finally it was placed in ethanol for 12 hours to remove organic contaminants from the surface. Before the trapping experiments, the chip was removed from the ethanol bath, immersed into toluene and dried with nitrogen flow. After this cleaning procedure the device was transferred into the chemical cell that was filled with a toluene solution containing the networks in low concentration (NR concentration around $10^{-9}$ M). A Suess probe station was employed to contact the two electrodes and a lock-in amplifier (SR7265, AMETEK Inc. USA) provided the AC voltage. The parameters of the trapping process (voltage amplitude, frequency, and trapping time) were computer controlled via a LABVIEW-based software interface. Fig. 2 shows a schematic drawing of the trapping mechanism. Before trapping, the gaps were checked to ensure that no electrical shorts or leakage currents were detectable.

The conductive properties of the trapped nanorod networks were measured with a Keithley 2612 source-meter in a two-probe configuration. All I-V curves were taken in air and at room temperature. For the photoelectrical properties, a diode laser emitting at 473 nm was used to irradiate the device with a power of 10 mW on a spot size with approximately 5 mm diameter. After the first set of measurements, the chip was annealed at 250 °C for 15 minutes under $N_2$ atmosphere, and the electrical properties were recorded again. All the scanning electron microscopy (SEM) imaging for this work was carried out with the Raith 150/2 system, which is a field-emission instrument with fairly high resolution.

## Results and discussion

A typical TEM image of the as-made networks is reported in Fig. 3, clearly showing chain-like nanostructures. In order to control the quantity of trapped networks, we regulated the voltage amplitude, frequency, and trapping time, as they proved to be the crucial parameters in that respect. Best results were obtained with a frequency of 1 kHz, both significantly higher and lower frequency values resulted in less efficient trapping. Increased trapping time resulted in a higher density of networks in-between the electrodes gap area. The most sensitive parameter was the voltage amplitude. Below a voltage amplitude of 3 $V_{RMS}$, no trapping of networks was observed. Above this threshold, the quantity of trapped networks increased with increasing voltage, as demonstrated in Fig. 4 a-b. Voltage amplitudes exceeding 8 $V_{RMS}$ led to damage of the electrodes. Our best set of parameters for trapping the CdSe-Au networks was a frequency of 1 kHz, a voltage amplitude of 3 $V_{RMS}$ and a trapping time of 60 s. With these parameters we obtained ~85% of our 30 pairs of electrode linked via small-scale networks. A representative result is shown in Fig. 4b, where the networks bridged the electrode gap and formed a conductive electrical pathway.

We note that the approach of electrode fabrication followed by dielectrophoresis for nanoparticle positioning has a number of advantages over other methods where the nanoparticles are deposited before electrode fabrication and then selectively contacted by an overlayer EBL process. In the dielectrophoresis approach, (i) the lithography process is much easier because it can be done on a clean wafer surface and there is no need for alignment (like when pre-deposited nanoparticles are in use), therefore resulting in a much higher throughput. Then, (ii) the nanoparticles do not suffer a hard post-deposition treatment, like PMMA coating, exposure to acetone etc., like it is the case in overlayer EBL processing.

Typical I-V characteristics of trapped networks prior to annealing are plotted in Fig. 5, showing reproducible and stable non-linear conductivity, but no photo-induced current. The I-V curves measured before annealing can be well described by a simple exponential model, where $I \sim I_0 \exp(|V|/V_0)$ (see Fig. 5), yielding a characteristic voltage of $V_0 = 0.76$ V.

After annealing the conductivity of the networks was significantly reduced, as shown in Fig. 6, but interestingly illumination of the device with blue laser light resulted in photo-induced current.

To understand the effect of the thermal annealing on the CdSe-Au network structure in more detail, we have applied a similar annealing process to networks that were deposited onto carbon-coated TEM grids. TEM images before and after thermal annealing are shown in Fig. 7, where we clearly observe not only Au domains that link the nanorod tips, but also many additional Au domains that decorated the lateral facets of the nanorods. The conductance of such networks was most likely mediated by thermally activated carriers and hopping charge transport in between localized state. Consequently, transport appears not to be sensitive to the semiconductor material and light irradiation. After annealing (Fig. 7b) the Au had migrated to the nanorod tips by ripening process as reported by Figuerola et al. (Figuerola et al. 2010), which is clearly evidenced by the larger Au domains that link the nanorod tips. Current transport in networks as in Fig. 7b should be mediated by chains of a sequence of Au domain/CdSe nanorod/Au domain and so on.

In this case the semiconductor nanorod material has a clear impact on the conductivity, which should lead to photo-induced current as observed in Fig. 6. Furthermore, the dark current in such a network could be limited by a series of Schottky barriers which should reduce significantly the conductivity in comparison to the hopping transport between the metal domains (Fig. 5) ( Subannajui et al. 2008), in good agreement with our experimental observations.

## Conclusion

We have demonstrated that CdSe-Au networks fabricated by colloidal synthesis can be controllably assembled in between bow-tie electrodes with ~100 nm separation by using a dielectrophoresis process. We measured the conductance of small-scale networks composed of CdSe nanorods linked via Au domains. Thermal annealing of our devices resulted in Au migration to the nanorod tips. This in turn led to a conductive path composed of quasi-linear chains of semiconductor nanorods and Au domains, showing photo-induced current as compared to the very low conductivity in the dark.

## Acknowledgement

This research was supported by the European Commission through the Marie-Curie ToK project NANOTAIL (MTKD-CT-2006-042459).

**Figure Captions:**

**Fig.1** Schematic drawing of nanorods (NRs), (a) individual NRs covered by surfactants, (b) trapped in-between a pair of electrodes, a high resistance state, (c) assembled by a nanowelding process into networks, and (d) networks trapped in a nanoscale gap, where the linking of the NRs via gold domains largely reduces the junction resistance.

**Fig. 2** Schematic drawing of the electronic circuit for the positioning of CdSe-Au networks in-between electrodes by dielectrophoresis.

**Fig. 3** TEM image of as-synthesized CdSe-Au networks; inset: a higher TEM magnification from the same ensemble.

**Fig. 4** SEM images of trapped CdSe-Au networks with trapping parameters of 1 kHz, 60 s, 5 $V_{RMS}$ (a), and 1 kHz, 60 s, and 3 $V_{RMS}$ (b).

**Fig. 5** Room temperature I-V curves of CdSe-Au networks with (open squares) and without (open circles) laser light illumination before thermal annealing. The solid line shows the result of the exponential fit to the data as described in text.

**Fig. 6** Room temperature I-V curves of CdSe-Au networks with (open squares) and without (open circles) laser light illumination after thermal annealing.

**Fig. 7** Detailed TEM images of (a) as-synthesized CdSe-Au nanorod networks showing also that small Au-domains decorate the surface of nanorods, (b) of post-annealed CdSe-Au networks, from the same batch, where metal domains are located only at the nanorod tips.

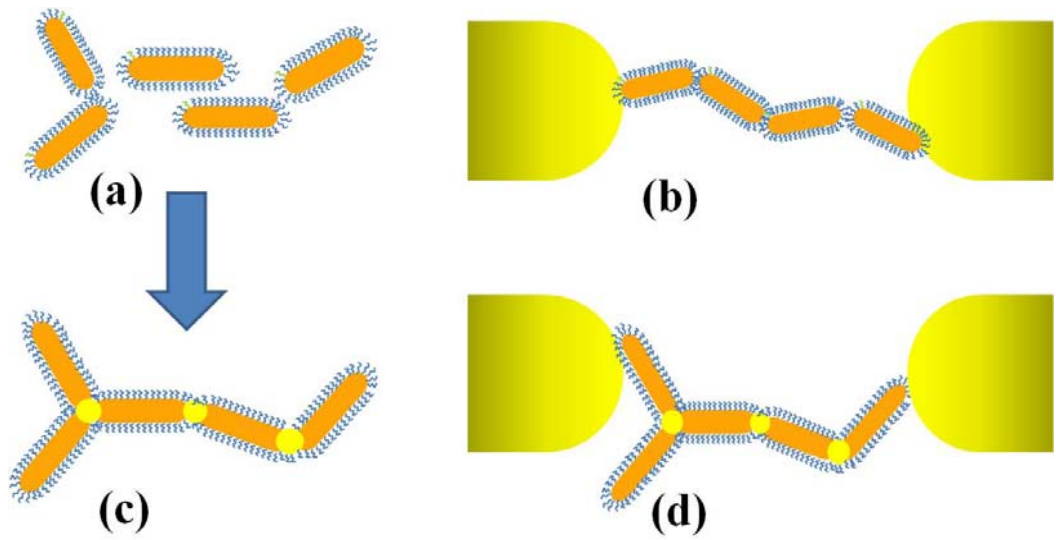

Figure 1

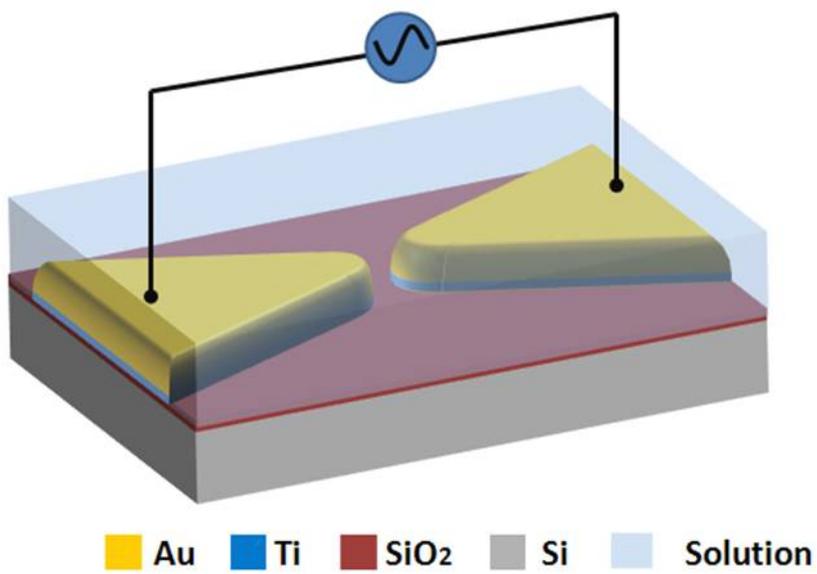

Figure 2

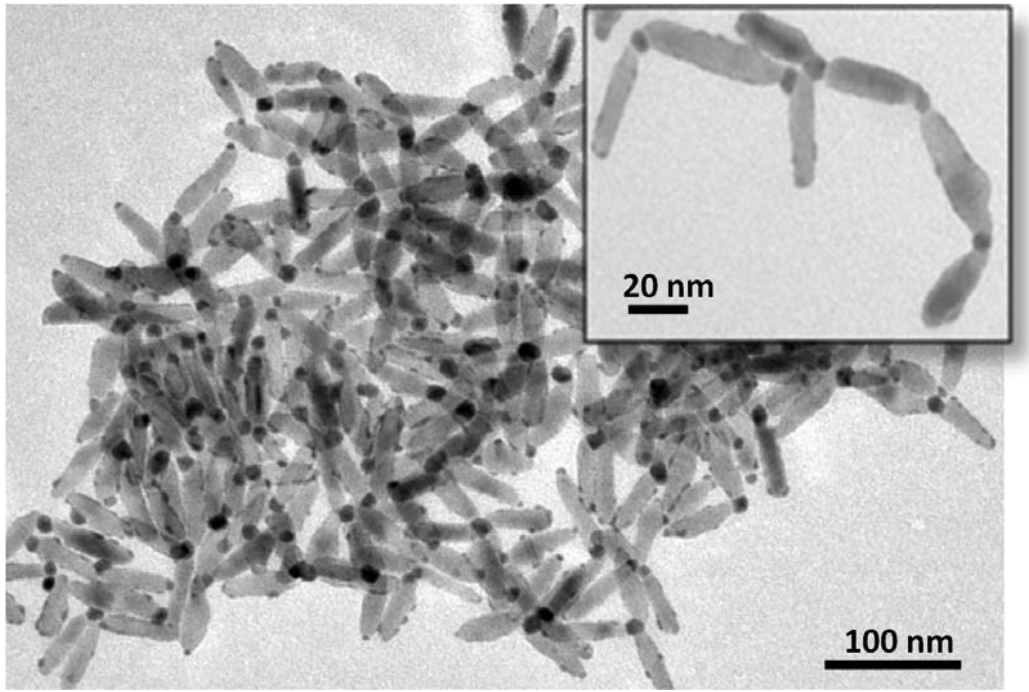

Figure 3

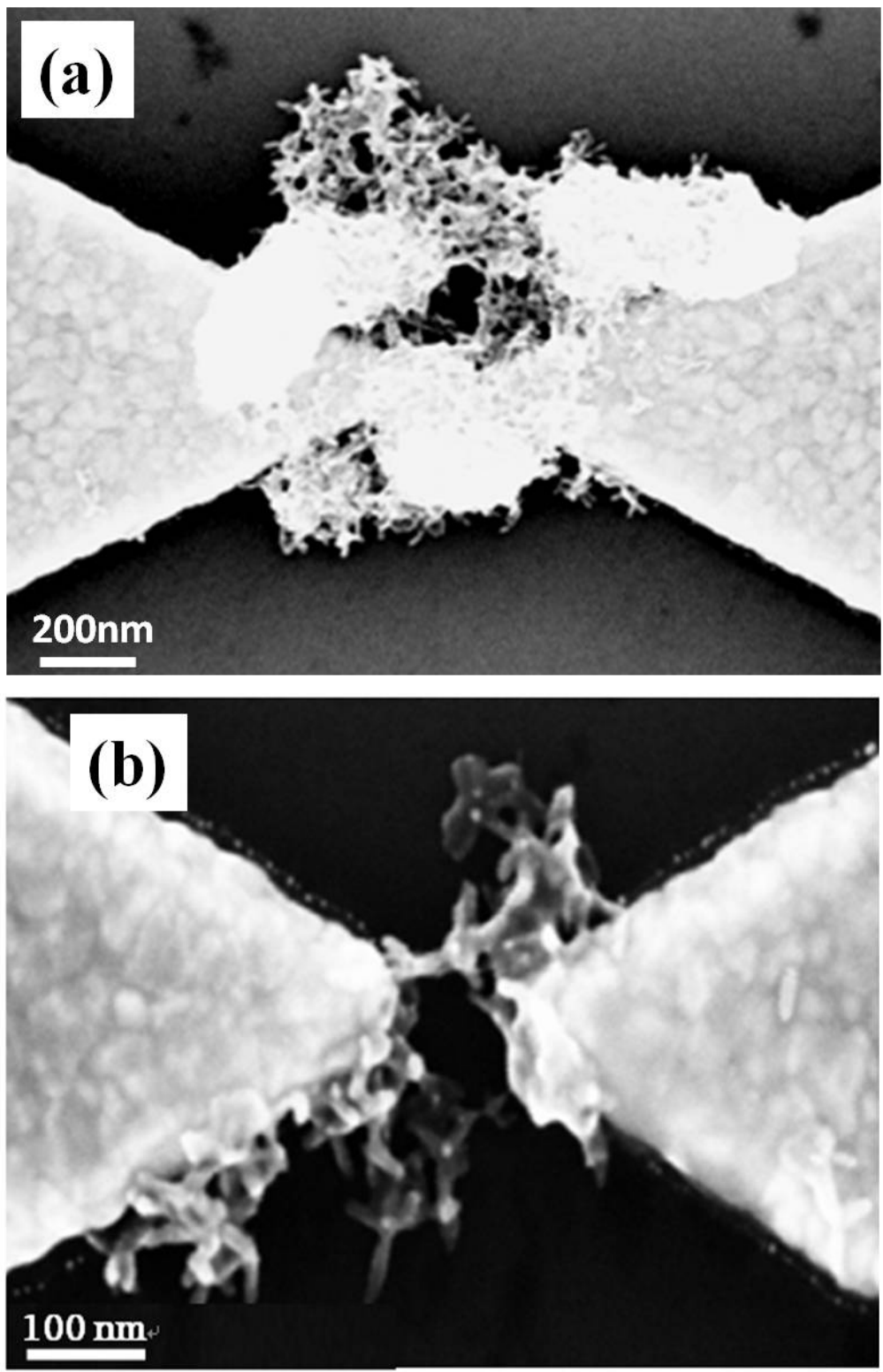

Figure 4

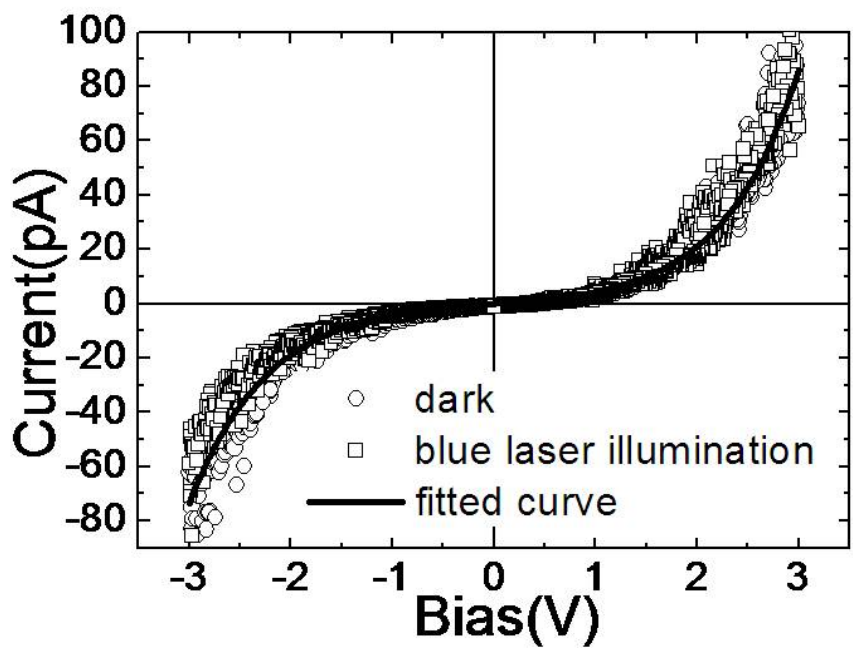

Figure 5

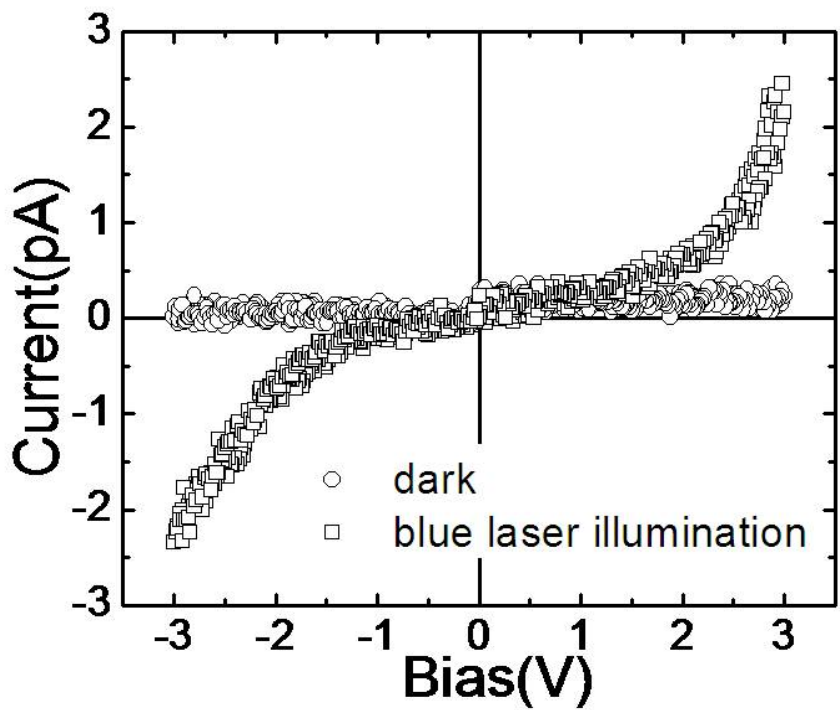

Figure 6

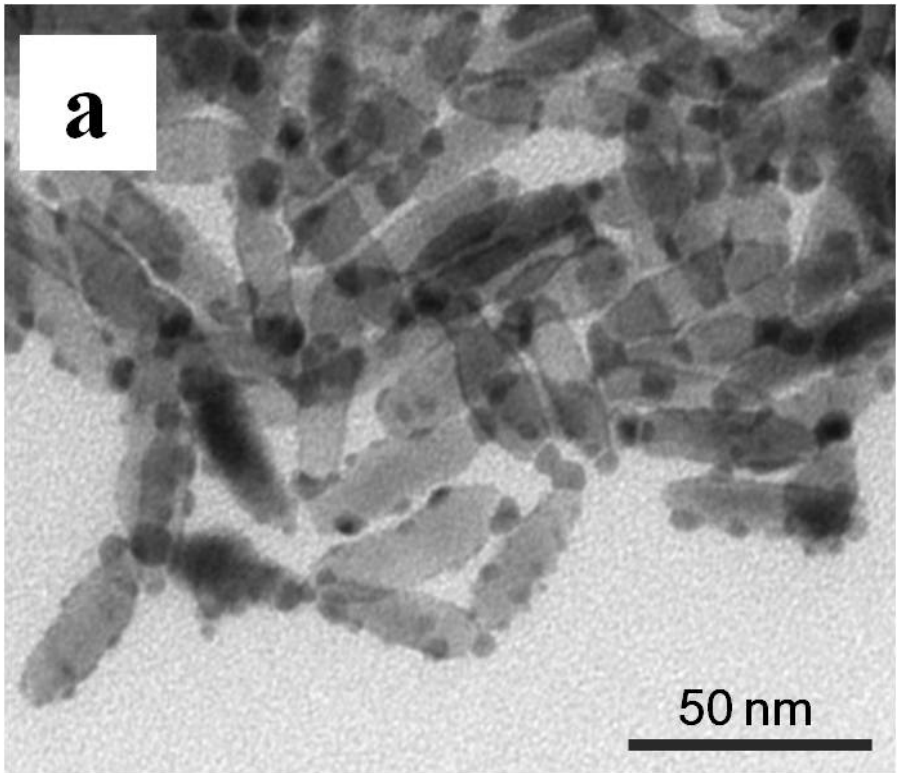

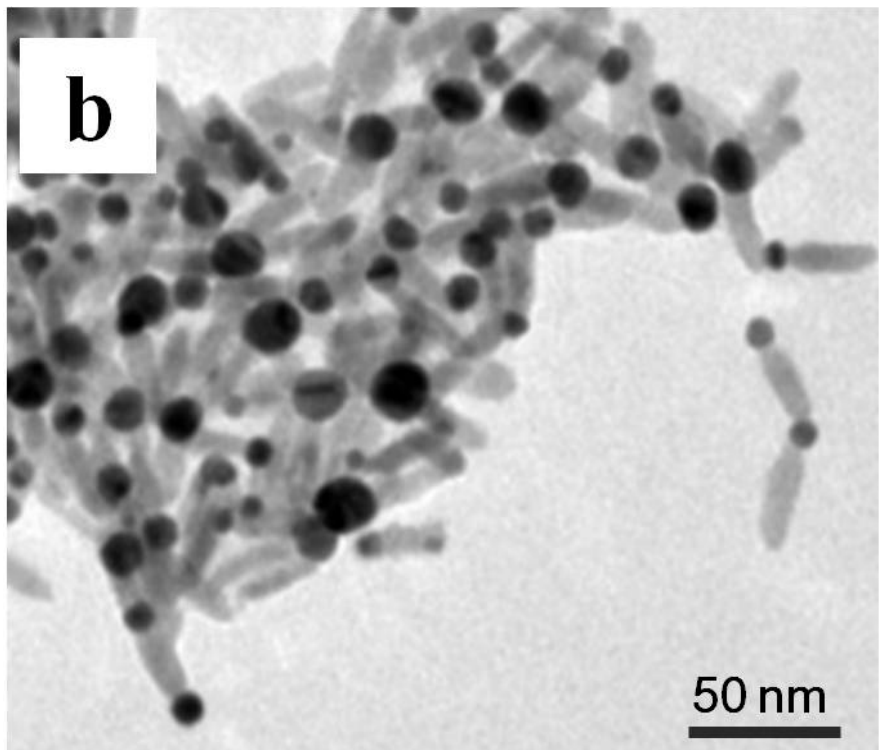

Figure 7